# SURFACE DRIVEN MN-DOPING OF GE QUANTUM DOTS - MN-INTERACTION WITH THE GE QD {105} FACET AND THE WETTING LAYER

C.A. Nolph, J.K. Kassim, J.A. Floro, and P. Reinke

Department of Materials Science and Engineering, University of Virginia, 395 McCormick Road, Charlottesville, VA 22901, U.S.A.

**ABSTRACT**

The interaction of Mn with Ge quantum dots (QD), which are bounded by {105} facets, and the strained Ge wetting layer (WL), terminated by a (001) surface, is investigated with scanning tunneling microscopy (STM). These surfaces constitute the growth surfaces in the growth of Mn-doped QDs. Mn is deposited on the Ge QD and WL surface in sub-monolayer concentrations, and subsequently annealed up to temperature of 400º C. The changes in bonding and surface topography were measured with STM during the annealing process. Mn forms flat islands on the Ge {105} facet, whose shape and position is guided by the rebonded step reconstruction of the facet. The voltage-dependent STM images indicate a hybridization between the Mn-d band and the empty states of the Ge{105} reconstruction. A statistical analysis of Mn-island shapes and position on the QD yields a slight preference for edge positions, whereas the QD strain field does not impact Mn-island position. However, the formation of ultra-small Mn-clusters dominates on the Ge(001) WL, which is in contrast to Mn-interaction with unstrained Ge(001) surfaces. Annealing to T<160º C leaves the Mn-clusters on the WL unchanged, while the Mn-islands on the Ge{105} facet undergo first a ripening process, followed by a volume gain which can be attributed to the onset of intermixing with Ge. This development is supported by the statistical analysis of island volume, size and size distribution. Increasing the annealing temperature to 220º and finally 375º C leads to a rapid increase in the Mn-surface diffusion as evidenced by the formation of larger, nanometer size clusters, which are identified as germanide $Mn_5Ge_3$ from a mass balance analysis. This reaction is accompanied by the disappearance of the original Mn-surface structures and de-wetting of Mn is complete. This study unravels the details of Mn-Ge interactions, and demonstrates the role of surface diffusion as a determinant in the growth of Mn-doped Ge materials. Surface doping of Ge-nanostructures at lower temperatures could provide a pathway to control magnetism in the Mn-Ge system.

**INTRODUCTION**

The scaling down of electronic devices continues to progress and spintronics offer solutions to key problems inherent to this development. In spintronics the spin degree of freedom of the electron is used to act as switch or store information, and spintronics memory as well as logic components have been developed [1,2]. One advantage in using spintronics-based device structures is the considerably smaller thermal load as compared to charge-based devices, which circumvents one of the most critical issues in current device architecture.

A fundamental component of spintronics is the development of dilute magnetic semiconductors (DMS) where magnetic dopants are incorporated into a semiconductor matrix and carriers mediate ferromagnetic order [3–5]. Much work has been done on III-V based DMS thin films and quantum dots [4–12] and the field is quite mature compared to the study of Mn-doped group IV semiconductors. Group IV based DMS have gained more attention in recent years partially due to predictions of higher Curie temperatures and their unique compatibility with the current charge-based Si-devices, and long spin lifetimes in Si [13].

Mn-doped Ge has especially garnered attention since the first reports of ferromagnetism in $Mn_xGe_{1-x}$ thin films by Park et al. [14], which included experimental evidence of Curie temperatures ($T_C$) up to 116 K, and a theoretical prediction of above room temperature $T_C$. This study inspired research in this area, and several reports of room temperature ferromagnetism or DMS behavior in Mn-doped Ge and Ge thin films followed [14–22]. However, the complexity of the Mn-Ge materials system proved to be a challenge defined by, first, the low solubility of Mn in Ge requiring non-equilibrium growth methods, and, second, the formation of second phases (germanides and silicides), which can be itinerant ferromagnets [23]. These ferromagnetic germanides are not conducive to the fabrication of DMS thin films and nanostructures. Germanide and presumably also silicide formation are particularly problematic in the growth of Mn-doped Ge quantum dots (QD) where a temperature of at least 400º C is required to trigger the nucleation of the QDs in the Stranski-Krastanov growth on a Si(001) substrate. This temperature regime favors at the same time the formation of secondary (Ge-Mn, Si-Mn) or even ternary phases (Ge-Si-Mn). The body of work from recent years illustrates that the magnetic signature of $Mn_xGe_{1-x}$ depends on a complex interplay between spatial distribution of Mn within the Ge matrix, and Mn-rich and poor regions, local bonding around Mn [substitutional or interstitial in Ge matrix, germanide], and the Mn concentration [14–16,18–20,24–26].

The magnetic doping of QDs is predicted to increase $T_C$ in DMS materials since quantum confinement can contribute to stronger hole-mediated coupling between the spin-centers defined by the Mn-dopant atoms [27,28]. The synthesis of Mn-doped Ge quantum dots with Curie temperatures above room temperature has recently been demonstrated by Xiu et al. [19,20]. These QDs were made by co-deposition of Ge and Mn on Si(001) within the typical Ge QD growth parameter window, and it was even shown that equally high $T_C$ can be achieved with Fe doping in the same manner [29]. The Mn doped Ge QDs exhibited segregation of Mn-rich regions below Ge QD dome islands, which illustrates a delicate interplay between Mn incorporation in the Ge matrix. The growth of Mn-doped QDs using the co-deposition process is also the subject of a recent study by Kassim et al. [25], which yielded new, and somewhat contradictory results on the relation between Mn-bonding, secondary phase formation and magnetism.

In order to control the Mn-doping of Ge QDs, and ultimately gain the ability to tailor its magnetic properties, we need to develop a better understanding of the growth process at the atomic scale. We therefore investigate the interaction and bonding between Mn and the two growth surfaces: the strained Ge(001) wetting layer (WL), and the Ge{105} facets, which are the boundary planes of Ge-QD hut clusters. We investigated the behavior of room-temperature deposited Mn on already-formed Ge QDs, and observed the transformation of the Mn distribution as a function of temperature with scanning tunneling microscopy (STM). This afforded insight into the mechanisms driving secondary phase formation. The surface-driven method gives unprecedented control of Mn-incorporation, and we envisage in the future the fabrication of embedded magnetic nanostructures in analogy to atomic scale device structures embedded in Ge as demonstrated by Scapucci et al. [30,31].

**EXPERIMENTAL**

The entire experiment cycle consisting of preparation, annealing and characterization, was performed in an Omicron Nanotechnology Variable Temperature SPM system with a base pressure of 8 x $10^{-11}$ mbar. The Si samples (B doped: 1.2 x $10^{17}$ – 1.5 x $10^{17}$ cm$^{-3}$) were annealed at approximately 550° C for 10 hours or longer until the system pressure had recovered back to the base pressure. Samples were then repeatedly flashed to progressively higher temperatures up to 1250° C, followed by a cool-down period of one hour [32,33]. Ge quantum dots were grown on Si substrates at a temperature of approximately 450° C at a Ge rate of 0.005 Å/s using a Veeco effusion cell. Mn was evaporated from a Mo-crucible using a home built electron beam evaporation source at a rate of 0.025 Å/s. The Mn-coverage ($\Theta_{Mn}$) was in the sub-monolayer regime, 0.1 – 0.2 ML.

Imaging conditions with gap voltages ranging from -2.0 V to +2.0 V were used with a 0.03 nA current set point, and imaging conditions are given in the respective figure captions. Many of the STM images were measured at elevated temperatures with the use of an Omicron Nanotechnology STM stage power supply to heat the samples by direct current heating. In addition the power supply applies a compensation voltage to offset the voltage drop across the length of the rectangular samples to obtain the correct imaging gap voltage. The heating to a given temperature was fast - several minutes until the temperature is stabilized, subsequently up to an hour before thermal drift was minimized during imaging. Measurements were then taken at the elevated temperature for on average one hour, after which the temperature was increased to the next setpoint. The surface was stable during the measurement window at a given temperature. The temperature measurement has a relatively large error of about ± 20º C due to the inherent difficulties in measuring sample temperatures on small Si samples below 500º C. A detailed description of the calibration procedure is given in [34].

Image analysis was performed using ImageJ [35], WSxM [36] and Gwyddion [37] software packages. Mn structures on Ge QDs were evaluated by their location on the QD, largest in-plane length, height, area and volume. To differentiate the location of the Mn structures on the QDs, the QD was divided into two regions: the facet region and the region extending approximately 1 nm beyond the edge where two facets meet called the edge region. The largest in-plane length and height were measured using line scans across the Mn structure. The area of the Mn structures were measured by first isolating the them from the background using n$^{th}$-order polynomial flattening processes (*n* was adjusted as needed to achieve optimum results where the Mn structures were visually separated from the surroundings with minimal distortion). Following the background flattening, a binary mask created by height thresholding generated an image containing only the Mn structures, and the total area occupied by this binary mask was measured. Similarly, the volume of the Mn structures was found by generating a mask including only the Mn structures and using a built-in function of Gwyddion to report the volume of the generated mask. The masked images were compared visually with the original images to ascertain correct thresholding.

**RESULTS**

**(A) Mn Adsorption at Room Temperature**

The results of our study are presented by first establishing the starting surfaces for the room temperature Mn deposition which are the Ge(001) wetting layer, and the Ge(105) facets of hut-shaped Ge QDs. The Ge surfaces prior to the deposition of Mn are shown in Fig. 1(a) for the Ge(001) wetting layer and Fig. 2(a) for the Ge{105} QD facets. Typical STM images of Mn islands on the Ge wetting layer and Ge QDs are shown in Fig. 1(b) and Fig. 2(c,d), respectively, for the room temperature deposition.

The Ge wetting layer prior to Mn deposition is shown in Fig. 1(a) and consists of regions of (2x1) and c(4x2) reconstruction surrounded by DVLs (dimer vacancy lines), which divide the surface into tile-like regions [38–40]. The reconstruction and deep grooves of the DVL lines are easily recognized in the line scan included in Fig. 1(c). After Mn deposition, the most notable change in the surface structure is a roughening of the wetting layer surface as shown in Fig. 1(b). The linescans include a typical profile across Mn adatoms/islands on the Ge wetting layer. The Mn islands have an apparent height with respect to the Ge-surface of 0.25 – 0.35 nm, but not all Mn-islands are as well-defined as the one shown in the figure. The measurement of size and exact position of these Mn-islands is unreliable since the island perimeter cannot be discerned unambiguously due to the underlying roughening of the wetting layer. A rough estimate yields an average inter-cluster distance of about 5 nm, and no preference for positioning with respect to the DVLs. The surface roughening can be seen in the line scan on each side of the island, and is evident in the direct comparison to the clean Ge wetting layer surface. However, the long-range tile structure defined by the DVLs is preserved [39,40]. The deposition of Mn disrupts the reconstruction of the wetting layer, which is only retained in small segments of the surface, although some segments of the WL surface appear to exhibit a longer range ordering (see linescan Fig. 1(c)). This is in contrast to deposition on the bulk Ge(001) surface [22], where Mn forms either small clusters or moves into sub-surface sites at lower temperatures but does not alter the Ge-surface reconstruction. The Ge wetting layer is compressively strained [41], and the presence of Mn could lead to a surface relaxation, which can account for the observed roughening and loss of reconstruction.

The Ge{105} facets, which are the smooth side facets on the hut shaped quantum dots in Fig. 2, show the rebonded step (RS) reconstruction [42–45]. This reconstruction is formed by rebonding at the $S_B$ steps of the Ge(001) nanofacets to minimize the number of surface dangling bonds. The rebonded step structure is recognized by characteristic empty and filled state STM images, which are shown in Figs. 2 and 3. On the Ge{105} QD facets, the room temperature deposition of Mn yields circular or ellipsoidal, flat islands with an extension of at most a few nanometers. The Mn islands are flat and have an apparent height, which corresponds to 1-2 atoms, only a very few islands are higher. However, it is not possible to resolve individual Mn atoms within the islands. Approximately two thirds of them are ellipsoidal and their long axis is mostly (90%) in the direction parallel to the horseshoe-shaped rebonded dimer structure, which is at a 45º angle with respect to the [010] and [$50\bar{1}$] directions indicated on the QD {105} facet.

The islands were then evaluated based on their position on the Ge QDs. Two regions on the QD were distinguished: one consisting of the face of Ge{105} facets, and a second region consisting of the area surrounding the edges where the {105} facets meet. This separation differentiates the more reactive edge regions from the flat facets. The number of islands located in the region surrounding the edges and the facet faces were normalized to the respective available areas of each region. Seventeen QDs and 179 Mn islands were analyzed in this process, and the data are included in Fig. 4. The density of islands on the facet and edges is $0.017 \pm 0.003$ islands/nm$^2$ and $0.035 \pm 0.01$ islands/nm$^2$, respectively. The spatial distribution of Mn islands indicates that the room temperature mobility of Mn is appreciable, and allows transport of a sizeable percentage of Mn-adatoms across the facet leading to attachment and nucleation of islands at the more reactive edges. However, this process does not entirely suppress the nucleation on the facet, leading us to conclude that the diffusion length of Mn adatoms is in the range of facet dimensions but not significantly longer. This argument assumes an energetic preference for edge bonding sites as compared to facet bonding sites. The strain field in QDs, which extends from the base (maximum strain) to the apex (minimum strain) of the QD [46] as the lattice relaxes from the Si lattice parameter to the Ge lattice parameter, does not influence the spatial distribution of the Mn islands. No preference for island nucleation at either extreme of the strain field is observed, which might be attributable to the modulation of the QD bulk strain field by a tensile strain component introduced by the reconstruction [42].

The higher resolution images in Fig. 2, and the bias-dependent constant current images in Fig. 3, which cover empty and filled states, provide additional insight in the bonding between Mn-islands and the reconstructed {105} facet surface. The filled state image in Fig. 2(d) shows several islands and the reconstruction seamlessly surrounds the Mn-islands, except for a few missing Ge-atoms at the island perimeter (indicated by arrows), which are not correlated to defects in the original facet surface. The image is somewhat degraded compared to the filled state images prior to Mn deposition (Fig. 2b), but the preservation of the RS reconstruction is confirmed in Fig. 3, which includes a series of constant current images recorded with varying bias voltage. In the empty state images (Fig. 3(e-g)) the RS reconstruction is expressed as a row of dimers in agreement with the literature [42–45]. However, the perimeter of the Mn-islands in these empty state images is disturbed, and the Ge-reconstruction is interrupted. This distortion is most prominent in the 1.6 V empty state image (Fig. 3e), where the reconstruction is lost over at least a 2 nm perimeter around the island.

The RS reconstruction is characterized by a broader, featureless band between $E_F$ and -2.5 eV in the ODOS (occupied density of states), which was interpreted as a mostly delocalized surface state [43]. In the UDOS (unoccupied DOS) the RS has two distinct, localized and relatively intense states, which stem from the surface dimers at 1.0 and 1.5 eV. These states are expected to dominate the images taken at 1.6 V bias voltage as shown in Fig. 3e, and their relative contributions are diminished at higher bias voltages. The dramatic image modulation in the vicinity of the Mn-islands for the empty state images taken at 1.6 V indicated that Mn donates charge into the localized empty states of the dimers and thus disrupts the RS reconstruction. This can also be described as a hybridization or backbonding of the Mn-d electrons with the empty states of the Ge {105} surface p-band. The lack of atomic resolution within the islands is at the same time an indication of a delocalization of the Mn-d electrons.

The orientation of the majority of islands at the same time implies a preferred bonding configuration between Mn island and the RS reconstruction and the islands are guided by the reconstruction and appear "nestled" between the horseshoe-shaped dimer groups of the reconstruction (see Fig. 2 right hand side). The anisotropy of island shapes is often explained by an inherent anisotropy in the diffusion for ad-atoms; the Si diffusion on the Si(100)-(2x1) surface is a well-known example in this context [33]. Theoretical work on the diffusion of ad-atoms on the strained and unstrained Ge {105} surface has been limited to Ge and Si ad-atoms, but shows an isotropic diffusion in the [010] and [501] directions [47,48]. If we tentatively assume that the isotropic diffusion equally applies to Mn-adatoms, then the island shape is indeed determined by the directionality of bonding between Mn-islands and the RS reconstruction and not by an anisotropy in motion across the surface. This bonding scenario has to be energetically favorable as compared to Mn-Mn bonding, which would lead to cluster formation as it is seen on the Ge(001) WL surface.

**(B) Annealing of Mn-Surface Islands**

The evolution of Mn islands with temperature was observed in-situ (measuring on a hot sample) for the temperature ranges (I) room temperature to 150º C, (II) 340º C and 400° C, and (III) one sample measured at 220º C. This division was defined by experimental boundary conditions, and the maximum permissible measurement time without significant degradation of the tip. The Mn-islands were characterized with respect to island number density, apparent height, long-axis length and volume and each datapoint contains the values for 60-80 annealed islands on several QDs. The measurement time for each interval was about one hour and the Mn-islands are stable within the timeframe of the experiment.

We will first cover the results from annealing the Mn islands up to a temperature of 150° C, which are summarized in Fig. 4. The effect of annealing up to a temperature of 150°C on island density for islands located on the edges and the facet are given in Fig. 4(a). With increasing annealing temperature, the total number of islands (facet edge plus facet face) decreases by 25-30%. Fig. 4(b) contains the island size and island height as a function of temperature; all images used in this analysis were taken with the same bias voltage of -2 V. The change in average apparent height from 0.22 nm to 0.3 nm is substantial, at 60° C the island dimension shows an abrupt increase, and a similarly steep increase in island height is seen at 80° C. In both cases, length and height remain nearly constant after their initial abrupt increases; facet and edge islands showed the same behavior and are therefore not plotted separately.

The distribution of long-axis lengths is included in Fig. 4(c) for 20º C, immediately after the height increase at 80º C, and 150º C. The *maximum* of the distribution (the *most probable* island length) shifts from 2.1 to about 2.5 nm when moving from room temperature to 150º C, and the increase in *average* island length is driven by the loss of shorter islands positioned on the left hand side of the distribution, and gain in medium size islands, albeit at the loss of very few unusually long islands > 4.5 nm. The average island area remains approximately constant with 4.3 ±0.3 nm$^2$ throughout the entire annealing cycle, albeit the spread in island area is relatively large and can easily include changes in the overall shape of islands such as, for example, a loss in width to accommodate for the increase in average length by 2-3 interatomic distances. An analysis of exact islands shapes goes beyond the precision of our image analysis. The changes in the island length distribution are commensurate with a ripening process. The loss of very few unusually large islands >4.5 nm is likely not statistically significant and might be due to their overall scarcity - a much larger sample size is required to represent them correctly.

The increase in the height of the Mn islands, however, is indicative of a process, which goes beyond a mere ripening of the Mn-islands, and includes the incorporation of Ge in the Mn-islands. This hypothesis is tested by taking a closer look at the change in islands volume as a function of temperature, which is shown in Fig. 5. The average island volume [total volume for all islands/number of islands] and the total island volume per unit area [sum over all islands within a unit area] are summarized in Fig. 5a. The average island volume increases from 0.79 to 1.55 nm$^3$ and the total volume of islands per unit area increases from 0.017 to 0.028 nm$^3$/nm$^2$.

Fig. 6 summarizes the results obtained for higher annealing temperatures: (a1, a2) show the morphology after annealing to 220º C (0.2 ML Mn deposited at room temperature), and (b,c) illustrate the development for 340º C and 375º C (0.1 ML Mn deposited at room temperature). The latter sample was subsequently annealed to 405º C, the highest temperature in our experiments, and did not show any further changes in surface morphology as compared to the lower temperatures for this sample. All sample images were recorded during annealing. All images taken at temperatures of 220º C lack the shallow surface islands, which are characteristic for the lower deposition temperatures. The WL and QD surfaces have reverted to their original surface structure; the quality of the wetting layer is recovered and the increased roughness induced by Mn deposition is removed. At the same time new structures are observed on the surface, and they appear as spherical, small crystallites, which are positioned mostly at the bottom of QDs and are dispersed on the WL surface. In Fig. 6b and 6c these structures are pointed out with arrows, and in Fig. 6(c) one of the crystallites is included in a higher resolution image. These crystallites are tentatively interpreted as secondary phases, most likely germanides.

Fig. 7 shows a direct comparison between (a) the room temperature deposit, which corresponds to 0.2 ML Mn, and (b) the same sample after annealing to 220º C; images of the same surface after annealing are also included in Fig. 6 (a,b). The image segment recorded at 220º C depicts a region without QDs, which are rather widely spaced in this sample. The spherical clusters are on average 2-2.5 nm high, and between 7 and 10 nm in width. The image shows a slight temperature induced drift, which is taken into account in the measurement of the cluster diameter. The cluster volume is between 14 and 70 nm$^3$, although the measurement of cluster dimensions always somewhat overestimates the cluster diameter due to tip shape convolution with the sidewall. From our previous work we estimate the error in volume calculation due to this effect to be around 10-20%. The number of atoms in all clusters within the image frame is ~26,000, if we assume the density of the Mn solid (7.2 g/cm$^3$) or the density of Mn$_5$Ge$_3$ (hexagonal D8$_8$ structure 7.25 g/cm$^3$) [49]. A complete ML of Mn contains approximately 5·10$^{14}$ atoms/cm$^2$ which gives about 16,900 atoms for 0.2 ML [65% of total number of atoms in clusters] within the image area as source material for the round clusters observed after annealing. The formation of Mn-rich germanide clusters, for example Mn$_5$Ge$_3$, is in good agreement with our estimate, and it appears unlikely that a large fraction of Mn is incorporated in the WL or the QDs [23].

**DISCUSSION**

The Mn-Ge system exhibits a considerable variability in bonding and structure, and the transitions between them has been identified in the annealing experiments. The room temperature deposition of Mn, which is the initial configuration for all experiments, results in flat Mn islands on the Ge{105} QD surface. These islands are commensurate with a wetting of this surface, the interaction between the adatoms and the Ge-surface and thus interfacial energy is sufficient to overcome the cohesive energy gained in the formation of small Mn-clusters. However, this is in direct contrast to the formation of nanoclusters on the Ge(001) WL surface, which can be attributed to the differences between the two surface reconstructions and a favorable bonding situation on the QD facets. The WL surface reacts to the Mn-deposition with a structural relaxation, an increase in roughness, and only small pockets of a reconstruction are maintained. In contrast, sub-monolayer Mn deposition on bulk Ge(001) substrates leads to distinct absorption geometries with a preference for an interstitial bonding site in the surface reconstruction as described by Zeng et al. [22]. The comparison with unstrained Ge-surfaces indicates that the strain in the Ge(001) WL surface drives nanocluster formation, and is detrimental to Mn adatom incorporation in surface sites.

The surface islands on the Ge{105} QD remain relatively flat and therefore can still be considered as "wetting" the surface for T<200ºC, but undergo a shape transformation through lateral and height growth. The islands themselves remain rather diffuse in the STM images – the individual Mn adatoms cannot be resolved, which is indicative either of metallic islands, or a rapid motion of the Mn atoms within the island. Even mild annealing induces a ripening process, and the increase in island size, and loss of small islands from the distribution, is commensurate with an Ostwald ripening process. The ripening process continues slowly after this initial increase in lateral size, but is now accompanied by a step-like rapid increase in island height from an average height of 1-2 atomic steps to 3-4 steps. The initial interpretation of this height increase is that Mn-adatoms have now sufficient energy to overcome the inverse Ehrlich-Schwöbel barrier and jump upwards at the island perimeter thus increasing island height. However, if we assume that these islands contain only Mn atoms the total Mn-island volume shown in Fig. 5 (right hand axis of graph) should be constant throughout the annealing process due to mass conservation.

An increase in average island volume can be explained by ripening, and the broadening of the average volume distribution shown in Fig 5b is commensurate with this process. However, the nearly twofold increase in total island volume per unit area (right hand axis, Fig. 5a) is only possible if additional atoms are incorporated in the islands assuming the height modulation is interpreted as a predominantly topographic effect. The volume analysis has a relatively large uncertainty in its measurement, but the initial increase in *total island volume* is indeed observed for T>80º C; the same temperature at which the height increase is initiated. We suggest that the volume increase is achieved by intermixing with Ge-surface atoms and local alloying. In the bulk phases the solubility of Ge in Mn reaches several percent, while the solubility of Mn in Ge is negligible [23]. The phase diagram of the bulk phases, however, can substantially deviate from that of surface and nanostructures, and is cited here to illustrate the asymmetry in solubility for the Mn and Ge system. The STM images of these islands are very similar to the room temperature images (Fig. 1) and taken at a bias voltage of 2.0 V. A detailed bonding analysis with atomic resolution was elusive at the elevated imaging temperatures.–Alternatively, the height increase might be understood as being caused by a change in LDOS, which would then be the signature of a substantial change in bonding between Mn and Ge-surface. Independent of whether the major contributor to the apparent height change is a topography or density of states variation, it has to be initiated by a reaction or re-distribution between the Mn-islands and the Ge-surface. However, the progression of island morphology as a function of temperature, and the magnitude of the change in island height leads us to favor an interpretation of the STM images as being dominated by topography rather than changes in LDOS.

The intermixing and reaction between the Mn-islands and the Ge QD surface can be interpreted as the signature of early stages of germanide formation. The Mn-clusters on the WL appear unaffected in the temperature range up to 150º C, albeit the difficulties in a reliable analysis of cluster dimensions can easily mask a moderate ripening and a reaction at the cluster-Ge interface. The changes encountered in the temperature range above 150º C are much more dramatic, and are characterized by the formation of large spherical clusters or crystallites presumably made of germanides according to the mass balance analysis. The complete removal of any Mn-related signatures on the QD facets and the WL surface indicates that Mn-adatoms diffuse across appreciable distances to contribute to the germanide nucleation and growth. In order to nucleate the germanide clusters it is necessary to provide a supersaturation of both reactants, and to be able to overcome a material specific activation barrier. The details of the germanide nucleation process remain speculative at present, although the transport of Mn across the surface is presumably the critical bottleneck in the reaction. The average distance between Mn-surface structures is an expression of the surface diffusion length of Mn adatoms as a function of temperature: the average distance between the low temperature Mn-islands and nanoclusters is only several nm, while the distance between the sizeable Mn-clusters at 220º C is already several ten nanometers. The germanide on the other hand has to provide a sufficient driving force in terms of a chemical potential gradient to promote Mn-diffusion, break up the WL Mn-nanoclusters and QD Mn-islands, and mobilize Ge surface atoms. The latter process draws from a very large reservoir, and weakly bound surface atoms provide an easily accessible source of Ge. However, we do not see a local pit formation in the Ge WL around the larger clusters at 220º C, which indicates a relatively large capture radius for the Ge atoms.

The reaction of Mn with a Ge(001) and Ge(111) substrate via solid state epitaxy leads to the formation of large Ge-crystallites, and continuous $Mn_5Ge_3$ thin films, respectively. The reaction between a Mn-deposit and the Ge substrate is usually initiated at temperatures exceeding 300º C and high quality thin films on Ge(111) require around 500º C [21]. The lower temperature limit has not been studied in detail, and its numerical value is therefore not a well-confirmed boundary for the onset of germanide formation. These experiments overall support nanoscale germanide cluster formation through a solid state reaction, although 220º C is a relatively low temperature for initiation of this process.

We note that our system is actually a ternary system, since a large reservoir of Si is present just below the thin Ge WL, and the reactivity of Mn with Si is on par with that of Ge. We performed a closely related study, which used higher temperature (>450º C) co-deposition of Mn and Ge onto Si (001). We find, in fact, that silicide phases preferentially form over germanides [25]. In the present work, however, given the low temperature deposition of Mn onto Ge, and the moderate annealing temperatures, germanide formation is likely kinetically more facile, and a substantial intermixing of Si and Ge is not expected at least in the lower temperature regime.

We have now demonstrated the feasibility to control bonding of Mn-nanostructures on Ge QDs and the WL surface via a targeted annealing process. A set of three different Mn-nanostructures are identified: (1) Mn-clusters on WL, and Mn-islands on QDs, (2) intermixed Mn islands QD which likely coexist with Mn-clusters on the WL, and (3) germanide clusters. They are expected to provide different magnetic signatures: α-Mn is antiferromagnetic but several of the germanide phases are ferromagnetic; $Mn_5Ge_3$ for example, has a Curie temperature of 296 K. The detailed knowledge of the bonding configuration with respect to Mn will now enable us to target specific nanostructured materials and establish the feasibility of using them in a device environment.

**CONCLUSIONS**

The interaction of Mn, which is one of the most promising elements for magnetic doping, with the Ge(001) WL and Ge{105} QD surfaces was studied with STM. The two surfaces show marked differences: while Mn forms clearly discernible small clusters on the WL, Mn ad-atoms form shallow islands on the QD facet, which are guided by the RS reconstruction. The bonding between the islands and the RS reconstruction can be described as a back-bonding of d-electrons from Mn into the empty states of the surface states. It can be concluded that the differences in surface structure are instrumental in driving the substantial differences in Mn-surface bonding. The strain within the QD has no bearing on the Mn-island distribution, but the strain within the WL drives nano-cluster formation which is concluded from the comparison to Mn on unstrained Ge(001) surfaces.

Annealing of the Mn surface structures on the QD {105} facets leads to an intermixing or alloying with Ge-atoms, which is interpreted as the early stages of germanide formation. A critical temperature in the annealing sequence is around 220º C, where these metastable, intermixed islands break up, and considerably larger germanide clusters begin to form on the WL and at the base of the QDs. This can be described as a de-wetting process, which includes a chemical reaction with the surface. The dramatic increase of diffusion length from a few nm at room temperature to several ten nm at 220º C is an important contribution to drive germanide cluster nucleation and growth. We have not enough information at present to assess the role of energy barriers to nucleation, and the size of critical germanide nuclei. The de-wetting is complete and the original Ge surface structures are completely recovered above the critical temperature. The Mn-surface structures on the WL and QD likely express rather different magnetic signatures, where the clusters on the WL are antiferromagnetic Mn, and the islands on the QD show a ferromagnetic coupling reminiscent of ferromagnetism observed in germanides or Mn-doped Ge thin films.

**ACKNOWLEDGEMENTS**

The authors gratefully acknowledge the support of this work by NSF with the award number DMR-0907234, Division of Materials Research (Electronic and Photonic Materials).

**FIGURE CAPTIONS**

**Fig. 1.** STM images of the Ge(001) wetting layer (a) prior to Mn deposition and (b) after Mn deposition of 0.1 ML Mn. (c) Linescans from (a) and (b) were taken from representative areas on each image and are indicated by a bar within the image. Both images were obtained with a sample bias of -2.0 V.

**Fig. 2.** STM images of a single Ge QD (a) prior to Mn deposition and (b) image of one QD facet containing an overlay of the Ge{105}-(2x1) rebonded step (RS) reconstruction. A schematic of the RS reconstruction is included on the right hand side: the continuous and broken lines are the $S_A$ and $S_B$ steps, respectively, the black and open circles denote atoms located on different terraces, and the red dots (online: red, print: grey) are the atoms highly visible in the filled state images of the surface. The crystallographic directions with respect to the facet on a typical hut shaped QD are indicated in the second schematics. (C) shows an image of a second Ge QD recorded after Mn deposition (0.1 ML) with Mn islands on the facets and facet edges of the QD. (D) is a smaller segment of the QD facet along with an overlay of the Ge{105}-(2x1) rebonded step reconstruction. All images are filled state images obtained with a sample bias of -2.0V.

**Fig. 3.** Images of the same Mn islands before annealing as a function of bias voltage. The filled state images are recorded with (a) -14. V, (b) -1.6 V, (c) -1.8 V, and (d) -2.0 V for the filled state images, and (e) 1.6 V, (f) 1.8 V, and (g) 2.0 V for the empty state images.

**Fig. 4.** (a) Number density of Mn islands on Ge QDs facets and facet edges as a function of annealing temperature from room temperature to 150° C. (b) Mn island size (long-axis length, left axis) and Mn island height (right axis) as a function of annealing temperature from room temperature to 150°C. (C) distributions of Mn island size (long-axis length) at room temperature, 80º C and 150º C, and. (d) distributions of Mn island height at room temperature, 80º C, 150° C. The filled dots mark the most probable island size and volume, the light gray dot marks the average island size and volume as included in the graphs (a,b).

**Fig. 5.** (A) The average Mn island volume , and total island volume normalized to unit area as a function of temperature, and (b) distributions of island volume at room temperature and at 150° C.

**Fig. 6.** STM images of one sample at 220º C (a1, a2), and a second sample at temperatures of (b) 340º C, and (c) 375º C. All images were obtained with a sample bias of -2.0V.

**Fig. 7.** (a) QD and WL after room temperature deposition of 0.2 ML Mn at room temperature, and (b) at an annealing temperature of 220º C. This image is from the same sample as the QD images in Fig. 6 (a1, a2), and shows a sample region in between QDs. A slight thermal drift is superimposed on the image.

**REFERENCES**

[1]  A. H. MacDonald, P. Schiffer, and N. Samarth, Nat. Mat. **4**, 195 (2005).
[2]  I. Zutic, J. Fabian, and S. Das Sarma, Rev. Mod. Phys. **76**, 323 (2004).
[3]  T. Dietl, H. Ohno, F. Matsukura, J. Cibert, and D. Ferrand, Science **287**, 1019 (2000).
[4]  K. W. Edmonds, Curr. Opin. Solid St. M. **10**, 108 (2006).
[5]  T. Jungwirth, J. Sinova, J. Mašek, J. Kučera, and A. H. MacDonald, Rev. Mod. Phys. **78**, 809 (2006).
[6]  R. Abolfath, P. Hawrylak, and I. Žutić, Phys. Rev. Lett. **98**, (2007).
[7]  M. Holub, S. Chakrabarti, S. Fathpour, P. Bhattacharya, Y. Lei, and S. Ghosh, Appl. Phys. Lett. **85**, 973 (2004).
[8]  Z. Yu-Hong, Z. Jian-Hua, B. Jing-Feng, W. Wei-Zhu, J. Yang, W. Xiao-Guang, and X. Jian-Bai, Chinese Phys. Lett. **24**, 2118 (2007).
[9]  B. Beschoten, P. A. Crowell, I. Malajovich, D. D. Awschalom, F. Matsukura, A. Shen, and H. Ohno, Phys. Rev. Lett. **83**, 3073 (1999).
[10] F. Matsukura, H. Ohno, A. Shen, and Y. Sugawara, Phys. Rev. B **57**, R2037 (1998).
[11] K. Moon and P. Lyu, Eur. Phys. J. B **36**, 593 (2003).
[12] A. M. Nazmul, T. Amemiya, Y. Shuto, S. Sugahara, and M. Tanaka, Phys. Rev. Lett. **95**, 017201 (2005).
[13] I. Appelbaum, B. Huang, and D. J. Monsma, Nature **447**, 295 (2007).
[14] Y. D. Park, A. T. Hanbicki, S. C. Erwin, C. S. Hellberg, J. M. Sullivan, J. E. Mattson, T. F. Ambrose, A. Wilson, G. Spanos, and B. T. Jonker, Science **295**, 651 (2002).
[15] S. Guchhait, M. Jamil, A. Mehta, E. Arenholz, G. Lian, A. LiFatou, D. A. Ferrer, J. T. Markert, L. Colombo, and S. K. Banerjee, Phys. Rev. B **84**, 024432 (2011).
[16] M. Bolduc, C. Awo-Affouda, A. Stollenwerk, M. B. Huang, F. G. Ramos, G. Agnello, and V. P. LaBella, Phys. Rev. B **71**, 033302 (2005).
[17] N. Pinto, L. Morresi, M. Ficcadenti, R. Murri, F. D'Orazio, F. Lucari, L. Boarino, and G. Amato, Phys. Rev. B **72**, 165203 (2005).
[18] Y. Wang, J. Zou, Z. Zhao, X. Han, X. Zhou, and K. L. Wang, Appl. Phys. Lett. **92**, 101913 (2008).
[19] F. Xiu, Y. Wang, J. Kim, A. Hong, J. Tang, A. P. Jacob, J. Zou, and K. L. Wang, Nat. Mater. **9**, 337 (2010).
[20] F. Xiu, Y. Wang, J. Kim, P. Upadhyaya, Y. Zhou, X. Kou, W. Han, R. K. Kawakami, J. Zou, and K. L. Wang, ACS Nano **4**, 4948 (2010).
[21] C. Zeng, S. C. Erwin, L. C. Feldman, A. P. Li, R. Jin, Y. Song, J. R. Thompson, and H. H. Weitering, Appl. Phys. Lett. **83**, 5002 (2003).
[22] C. Zeng, Z. Zhang, K. van Benthem, M. F. Chisholm, and H. H. Weitering, Phys. Rev. Lett. **100**, 066101 (2008).
[23] A. B. Gokhale and R. Abbaschian, Bull. of Alloy Phase Diagrams **11**, 460 (1990).
[24] M. Jamet, A. Barksi, T. Devillers, V. Poydenot, R. Dujardin, P. Bayle-Guilemaud, J. Rothman, E. Belle-Amalric, A. Marty, J. Cibert, R. Mattana, and S. Tatarenko, Nature Materials **5**, 653 (2006).
[25] J. K. Kassim, C. A. Nolph, M. Jamet, S. Smith, P. Reinke, and J. Floro, Appl. Phys. Lett. submitted (2012).

[26] C. Zeng, W. Zhu, S. C. Erwin, Z. Zhang, and H. H. Weitering, Phys. Rev. B **70**, 205340 (2004).
[27] S. Sapra, D. D. Sarma, S. Sanvito, and N. A. Hill, Nano Lett. **2**, 605 (2002).
[28] J. Van Bree, P. M. Koenraad, and J. Fernandez-Rossier, Phys. Rev. B **78**, 165414 (2008).
[29] F. Xiu, Y. Wang, X. Kou, P. Upadhyaya, Y. Zhou, J. Zou, and K. Wang, J. Amer. Chem. Soc. **132**, 11425 (2010).
[30] G. Scappucci, G. Capellini, B. Johnston, W. M. Klesse, J. A. Miwa, and M. Y. Simmons, Nano Lett. **11**, 2272 (2011).
[31] G. Scappucci, G. Capellini, W. M. Klesse, and M. Y. Simmons, Nanotechnology **22**, 375203 (2011).
[32] K. Hata, T. Kimura, S. Ozawa, and H. Shigekawa, J. Vac. Sci. Technol. A **18**, 1933 (2000).
[33] B. S. Swartzentruber, Phys. Rev. Lett. **76**, 459 (1996).
[34] C. A. Nolph, K. R. Simov, H. Liu, and P. Reinke, J. Phys. Chem. C **114**, 19727 (2010).
[35] M. D. Abramoff, P. J. Magelhaes, and S. J. Ram, Biophotonics International **11**, 36 (2004).
[36] I. Horcas, R. Fernandez, J. M. Gómez-Rodríguez, J. Colchero, J. Gómez-Herrero, and A. M. Baro, Rev. Sci. Instrum. **78**, 013705 (2007).
[37] P. Klapetek and D. Nečas, *Gwyddion* (http://gwyddion.net/).
[38] U. Köhler, O. Jusko, B. Müller, M. H. Hoegen, and M. Pook, Ultramicroscopy **42–44, Part 1**, 832 (1992).
[39] F. Liu and M. G. Lagally, Phys. Rev. Lett. **76**, 3156 (1996).
[40] F. Wu, X. Chen, Z. Zhang, and M. Lagally, Physical Review Letters **74**, 574 (1995).
[41] A. Vailionis, B. Cho, P. Desjardins, D. G. Cahill, and J. E. Greene, Phys. Rev. Lett. **85**, 3672 (2000).
[42] Y. Fujikawa, K. Akiyama, T. Nagao, T. Sakurai, M. G. Lagally, T. Hashimoto, Y. Morikawa, and K. Terakura, Phys. Rev. Lett. **88**, 176101 (2002).
[43] Y. Fujikawa, T. Sakurai, and M. G. Lagally, Appl. Surf. Sci. **252**, 5244 (2006).
[44] T. Hashimoto, Y. Morikawa, Y. Fujikawa, T. Sakurai, M. G. Lagally, and K. Terakura, Surface Science **513**, L445 (2002).
[45] P. Raiteri, D. B. Migas, L. Miglio, A. Rastelli, and H. von Kaenel, Phys. Rev. Lett. **88**, 256103 (2002).
[46] W. Yu and A. Madhukar, Phys. Rev. Lett. **79**, 905 (1997).
[47] L. Huang, G.-H. Lu, F. Liu, and X. G. Gong, Surface Science **601**, 3067 (2007).
[48] F. Montalenti, D. B. Migas, F. Gamba, and L. Miglio, Phys. Rev. B **70**, 245315 (2004).
[49] H. Kim, G.-E. Jung, K. Yoon, K. H. Chung, and S.-J. Kahng, Surf. Sci. **602**, 481 (2008).

FIGURE 1

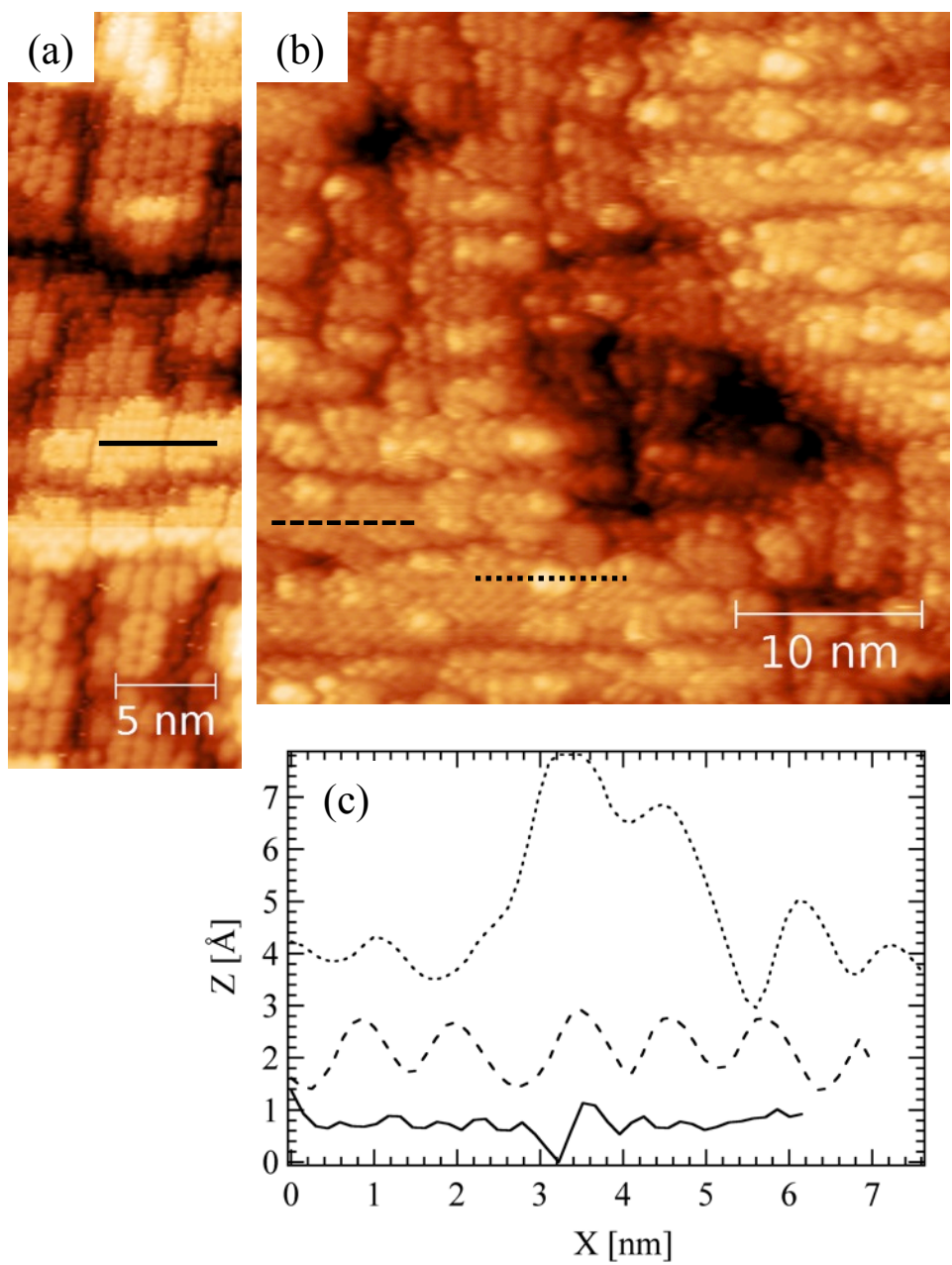

FIGURE 2

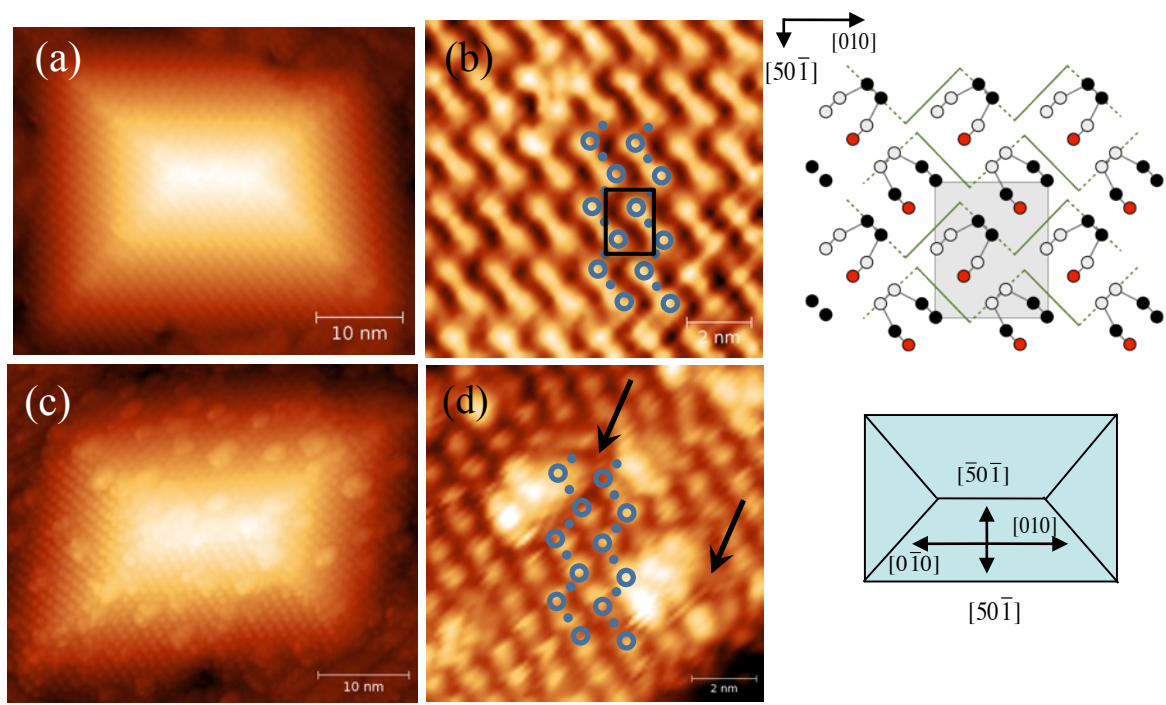

FIGURE 3

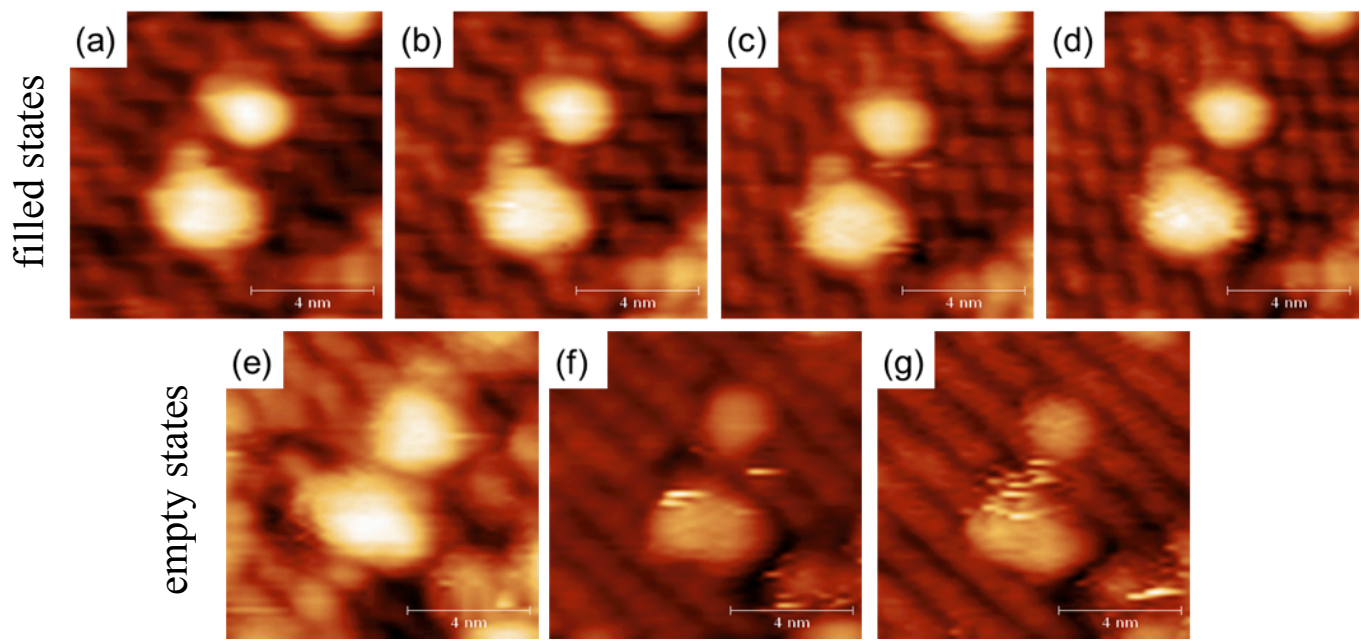

FIGURE 4

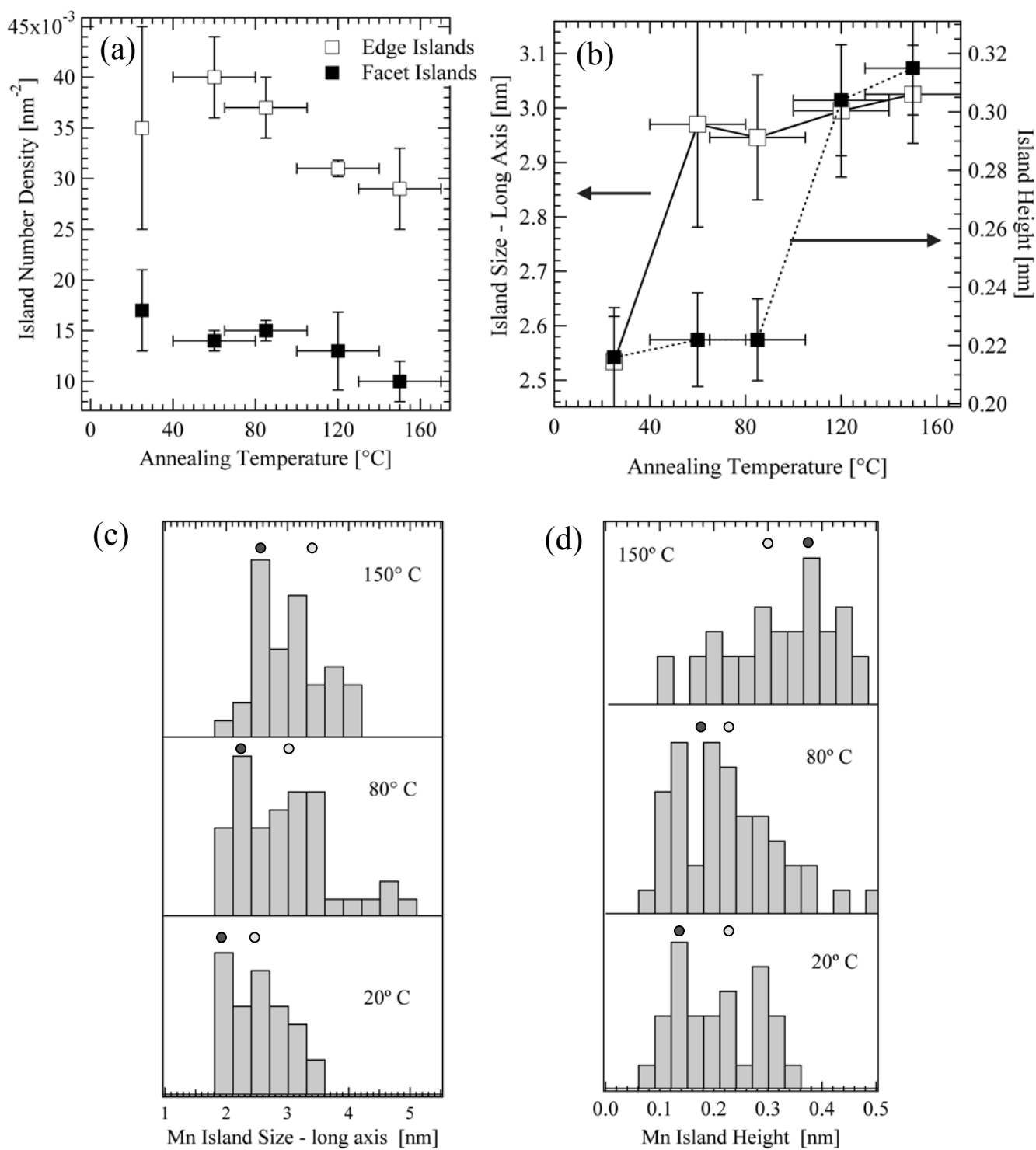

FIGURE 5

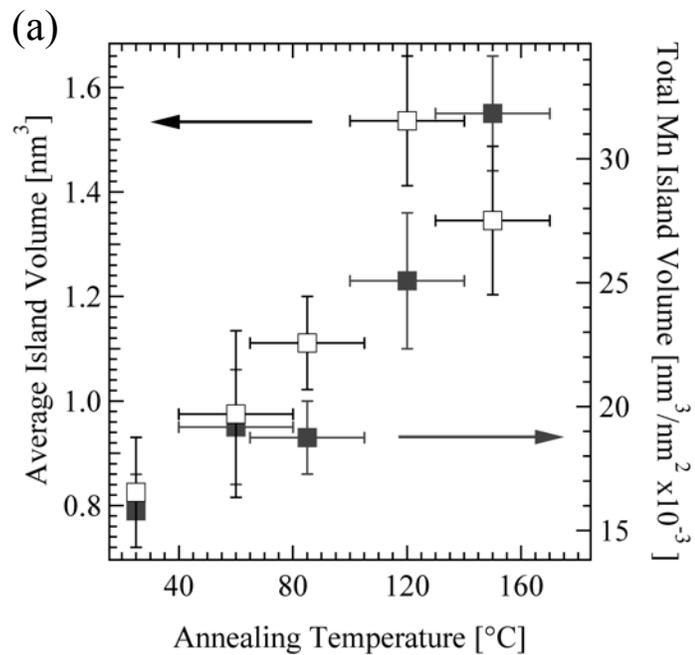

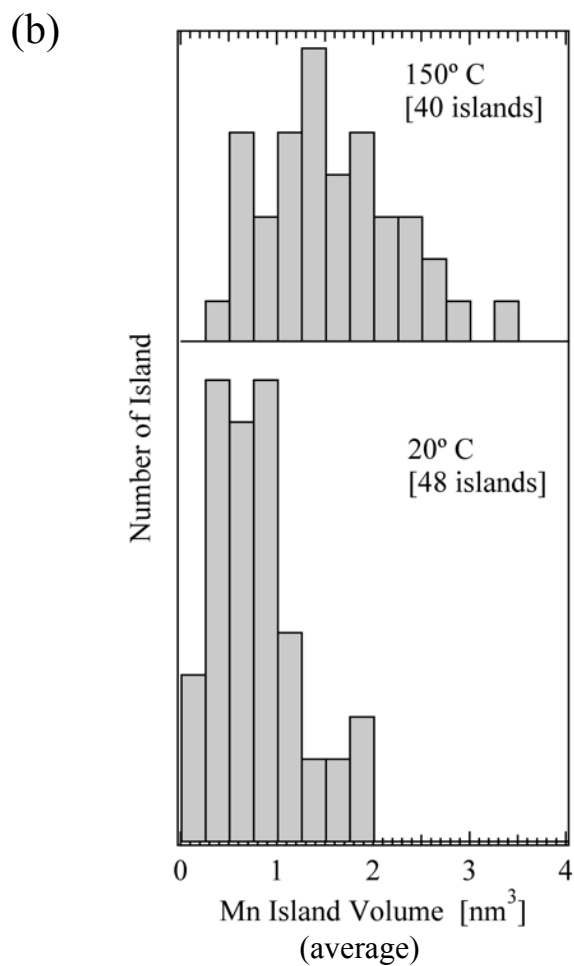

FIGURE 6

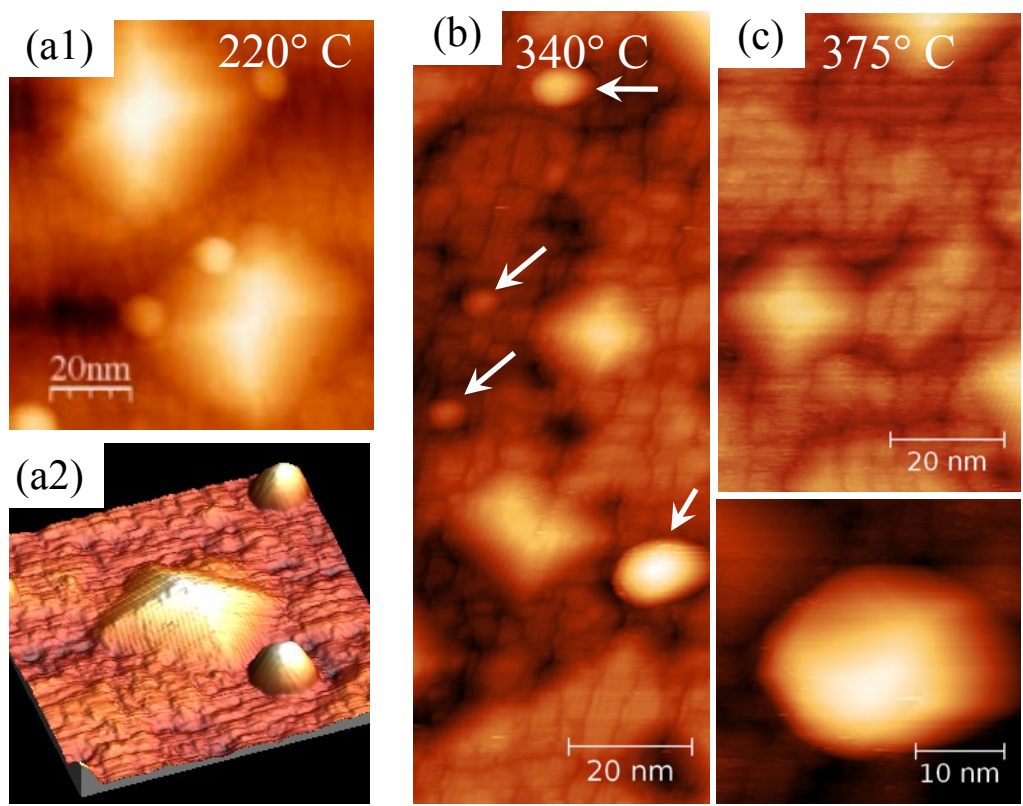

FIGURE 7

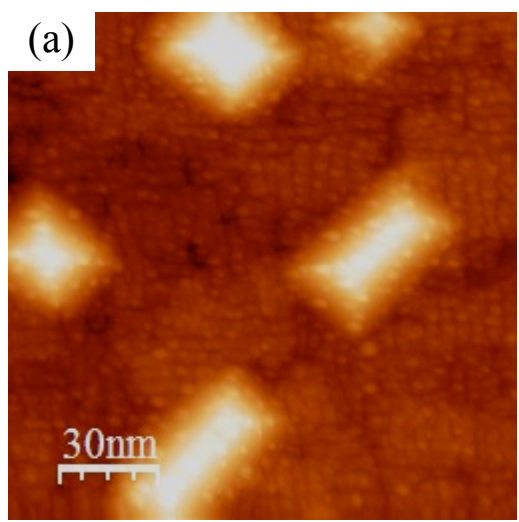 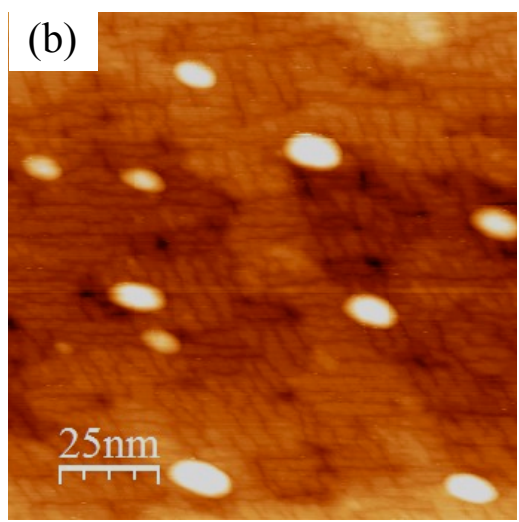